\documentclass[twocolumn]{svjour3}          
\usepackage[english]{babel}
\usepackage{amsmath,amssymb}

\DeclareMathOperator{\LE}{LE}
\DeclareMathOperator{\LCE}{LCE}

\begin{document}

\title{
  Invariance of
  Lyapunov exponents and Lyapunov dimension
  for regular and irregular linearizations
}
\author{N.V.~Kuznetsov \and T.A.~Alexeeva \and G.A.~Leonov}

 \institute{
 N.V. Kuznetsov
 \at
     Saint-Petersburg State University, Russia\\
     Department of Mathematical Information Technology,
     University of Jyv\"{a}skyl\"{a}, Jyv\"{a}skyl\"{a}, Finland \\
     \email{nkuznetsov239@gmail.com}
 \and
 T.A.~Alexeeva
 \at
     National Research University Higher School of Economics, Russia
 \and
 G.A.~Leonov
 \at
    Saint-Petersburg State University, Russia \\
    Institute of Problems of Mechanical Engineering RAS, Russia
 }
\date{Received: date / Accepted: date}
\maketitle

\begin{abstract}
Nowadays the Lyapunov exponents and Lyapunov dimension
have become so widespread and common
that they are often used without references to the rigorous definitions or pioneering works.
It may lead to a confusion since
there are at least two well-known definitions,
which are used in computations:
the upper bounds of the exponential growth rate of the norms of
linearized system solutions
(\emph{Lyapunov characteristic exponents}, LCEs)
and the upper bounds of the exponential growth rate of the singular
values of the fundamental matrix of linearized system (\emph{Lyapunov exponents}, LEs).
In this work the relation between Lyapunov exponents and Lyapunov characteristic exponents is discussed.
The invariance of Lyapunov exponents for regular and irregular linearizations
under the change of coordinates is demonstrated. 
\\
\end{abstract}

\section{Introduction}

Consider a continuous autonomous system
 \begin{equation} \label{sf}
 \begin{aligned}
 & \dot z=F(z),\quad z\in \,{\Bbb R}\,^n,
 \end{aligned}
 \end{equation}
where $F$ is a sufficiently smooth vector-function.
Suppose $z(t,x_0)$ is a solution of system (\ref{sf})
with the initial data $x_0 = z(0,x_0)$
uniformly bounded for $t \in [0,+\infty)$.
Consider the linearization of system (\ref{sf})
along the solution $z(t,x_0)$:
\begin{equation} \label{sfl}
 \dot x = J(t,x_0)x, \quad t \in [0,+\infty),
\end{equation}
where $J(t,x_0) = \{ \partial F_i(z)/ \partial z_j \}|_{z=z(t,x_0)}$
is ($n\times n$) Jacobi matrix.

Consider a fundamental matrix
\[
  X(t,x_0)=\big(x_1(t,x_0),...,x_n(t,x_0)\big),
\]
which consists of the linearly independent solutions \\ $\{x_i(t,x_0)\}_{1}^{n}$ of
linearized system \eqref{sfl}.
The fundamental matrix is often assumed to satisfy
the following condition: $X(0,x_0) = I_{n}$, where $I_{n}$
is a unit $(n\times n)$-matrix.

For time-varying linearization of nonlinear systems,
A.M.~Lyapunov introduced the so-called Lyapunov characteristic exponents (LCEs)
as the upper bounds of the exponential growth rates of solutions \cite{Lyapunov-1892}.

\begin{definition}\label{defLCE} \cite{Lyapunov-1892}\footnote{
In \cite{Lyapunov-1892} these values are defined
with the opposite sign and called \emph{characteristic exponents}.
}.
The Lyapunov characteristic exponents (LCEs) of matrix $X(t,x_0)$ are defined as
\[
 \LCE_i(x_0) =  \limsup\limits_{t \to +\infty} \frac{1}{t}\ln|x_i(t,x_0)|,
 \quad i=1,..,n.
\]
\end{definition}
For a scalar function $f(t,x_0) \neq 0$ and $t>0$
we introduce $\mathcal{X}(f(t,x_0))=\frac{1}{t}\ln|f(t,x_0)|$.
Then
\begin{equation}\label{xiprop}
  \mathcal{X}(|x_i(t,x_0)||x_j(t,x_0)|) = \mathcal{X}(|x_i(t,x_0)|)+ \mathcal{X}(|x_j(t,x_0)|)
\end{equation}
 and
$\LCE_i(t,x_0) =\mathcal{X}(x_i(t,x_0))$.

LCEs (often named the Lyapunov exponents or
Lyapunov exponents of fundamental matrix columns)
are used for the study of the existence of chaotic behavior in the theory of chaos
and the computation of dimension of chaotic attractors.
For invariant compact set of trajectories,
various coverages and their change along trajectories can be considered
in computing dimensions (see, e.g. \cite{BoichenkoLR-2005}).
If it is considered a coverage of attractor by small hypercubes
(see, e.g. survey \cite{GrassbergerP-1983}),
then LCEs characterize the exponential growth rates
of hypercube's edges lengths under the linearized map
(a unit hypercube is transformed by the fundamental matrix to a parallelotope,
the edges of which are the columns $x_i(t,x_0)$ of fundamental matrix,
and the volume is equal to $|\det X(t,x_0)|$).

Often in the dimension theory it is considered a coverage by balls
(see, e.g. definitions of the Hausdorff and fractal dimensions).
Here a unit ball $B$ is transformed into the ellipsoid $X(t,x_0)B$
and the exponential growth rates of its principal semiaxes lengths is considered.
The principal semiaxes of the ellipsoid $X(t,x_0)B$ are coincides
with the singular values of matrix $X(t,x_0)$:
$\sigma_i(t,x_0) = \sigma_i(X(t,x_0))$,
which are defined as the square roots of the eigenvalues
of matrix $X(t,x_0)^*X(t,x_0)$.
The exponential growth rates of the singular values are considered,
e.g. in \cite{Oseledec-1968}.

\begin{definition}\label{defLE}
The Lyapunov exponents (LEs) of matrix $X(t,x_0)$ are defined as
\[
   \LE_i(x_0) = \limsup\limits_{t \to +\infty}\mathcal{X}(\sigma_i(t,x_0)), \quad i=1,..,n.
\]
\end{definition}

In contrast to the stability theory,
where it is important to know only the largest LCE or LE,
in the chaos theory it is important to know all their possible values.
Therefore it is natural to consider the ordered sets LCEs and LEs.
For this purpose, considering the decreasing sequences
\[
  \LCE_i(t',x_0) = \LCE_i(X(t',x_0)) = \mathcal{X}({x_i}(t',x_0))
\]
\[
  \LE_i(t',x_0)= \LE_i(X(t',x_0))= \mathcal{X}(\sigma_i(t',x_0))
\]
for each $t=t'$
(called finite time LEs and LCEs, respectively),
one obtains the ordered (for all considered $t$)
sets of functions
\[
 \begin{aligned}
 & \LCE^o_1(t,x_0)\geq \LCE^o_2(t,x_0) \geq ... \geq \LCE^o_n(t,x_0), \\
 & \LE^o_1(t,x_0)\geq \LE^o_2(t,x_0) \geq ... \geq \LE^o_n(t,x_0),
 \end{aligned}
 \quad \forall t.
\]
Using, e.g, Courant-Fischer theorem \cite{HornJ-1994-book},
it is possible to show that the ordered $\LCE^{o}$s majorize the ordered $\LE^{o}$s: $\LE^o_i(t,x_0) \leq \LCE^o_i(t,x_0)$\footnote{
 For example,
 for the fundamental matrix
 \(
    X(t)=\left(
      \begin{array}{cc}
        1 & g(t)-g^{-1}(t) \\
        0 & 1 \\
      \end{array}
    \right)
 \) we have the following ordered values:
 $  \LCE_1^{o} =
  {\rm max}\big(\limsup\limits_{t \to +\infty}\mathcal{X}(g(t)),
  \limsup\limits_{t \to +\infty}\mathcal{X}(g^{-1}(t))\big),
  \LCE_2^{o} = 0$;
 $
  \LE_{1,2}^{o} = {\rm max, min}
  \big(
     \limsup\limits_{t \to +\infty}\mathcal{X}(g(t)),
     \limsup\limits_{t \to +\infty}\mathcal{X}(g^{-1}(t))
  \big).
 $ Remark that here $\mathcal{X}$ of the diagonal elements of $X(t)$
 do not coincide with $\LCE$s and $\LE$s.
}
and the largest $\LCE^{o}$ is equal to the largest $\LE^{o}$: $\LCE^{o}_1(t,x_0) =\LE^{o}_1(t,x_0)$.
For the sums of exponents the above fact has a simple geometric sense:
the volume of n-dimensional parallelotope (n-dimensional parallelepiped)
is less or equal to the volume of n-dimensional rectangular parallelotope with the same lengths of edges:
$|\det(X(t,x_0))| \leq |x_1(t,x_0)|\cdot\cdot\cdot |x_n(t,x_0)|$.
Therefore we have
\[
\begin{aligned}
  & \sum\limits_{i=1}^{n}\mathcal{X}(|\sigma_i(t,x_0)|)=\mathcal{X}(|\det(X(t,x_0))|) \leq \\
  & \leq
   \mathcal{X}(|x_1(t,x_0)|\cdot\cdot\cdot |x_n(t,x_0)|)=
  \sum\limits_{i=1}^{n}\mathcal{X}(|x_i(t,x_0)|).
\end{aligned}\]
If in the above definitions the limits exist (i.e. $\limsup\limits_{t \to +\infty}...$ is equal to $\liminf\limits_{t \to +\infty}...$
is equal to
$\lim\limits_{t \to +\infty}...$),
then, obviously,
it is sufficient to order the limit values $\LCE_i(x_0)$ and $\LE_i(x_0)$.

For a given point $x_0$ there is the essential question
on the existence of
$\lim\limits_{t \to +\infty}\LCE^o_i(t,x_0)$
and
$\lim\limits_{t \to +\infty}\LE^o_i(t,x_0)$.
In order to get rigorously the positive answer to these questions,
from a theoretical point of view, one may use
ergodic properties of the system:
according to the Oseledec theorem \cite{Oseledec-1968},
the limit exist for $x_0$ from a subset, which is the support
of an ergodic measure.
However, from a practical point of view,
the rigorous use of the above results is a challenging task
(e.g. even for the well-studied Lorenz system)
and hardly can be done effectively in the general case,
see, e.g. the corresponding discussions
in \cite{BarreiraS-2000},\cite[p.118]{ChaosBook}),\cite{OttY-2008},\cite{Young-2013}).


From a computational perspective, the essential justification of the existence
 of exact values of LCEs and LEs in numerical experiments may be the following:
 in the calculations with finite precision
 any bounded pseudo-trajectory $\widetilde{x}(t,x_0)$ has
 a point of self-intersection:
 $\exists t_1, t_2: \widetilde{x}(t_1,x_0) = \widetilde{x}(t_1+t_2,x_0)$.
 Then for sufficiently large $t\geq t_1$ the trajectory
 $\widetilde{x}(t,x_0)$ may be regarded as the periodic one.
 From a theoretical point of view,
 this fact relies on the shadowing theory,
 the closing lemma and its various generalizations
 (see, e.g. the surveys \cite{Mane-1984,Pilyugin-2011,Hertz-2013,Sambarino-2014}).

Lyapunov \cite{Lyapunov-1892} considered linearizations with boun\-ded coefficients
and introduced a special class of \emph{regular} linearizations:
for regular linearization the sum of LCEs equals to the lower bound
of the exponential growth rate of fundamental matrix determinant
(otherwise, the linearization is called irregular).
For example, the linearizations with constant and periodic coefficients are regular \cite{Lyapunov-1892}.
Any regular linearization has exact LCEs
(i.e. the $\lim\limits_{t \to +\infty}$ exists in the definition). 
Note that, in  general, the existence of exact LCEs does not imply
the regularity of linearization
(see, e.g., \cite{BylovVGN-1966,Barabanov-2005,LeonovK-2007}).
For regular linearizations the sign of the largest LCE defines stability/instability
of nonlinear system in a neighborhood of the considered solution
(see, e.g. theorems on stability by the first approximation
in the sense of Lyapunov and on instability in the sense of Krasovsky
and their various generalizations in survey \cite{LeonovK-2007}).
In the general case the sign of the largest LCE does not guarantee the stability
or instability (and, therefore, the positive largest LCE may not guarantee chaos)
since there are known the Perron effects of the largest LCE sign reversal
for irregular linearizations \cite{KuznetsovL-2001,LeonovK-2007,KuznetsovL-2005}.

Note that there are various essential generalizations of LCEs or LEs
(see, e.g. \cite{BylovVGN-1966,Pesin-1977,ConstantinFT-1985,Izobov-2012,CzornikNN-2013}).
See also some recent papers on LEs application
\cite{DeroinD-2015,Shevchenko-2014,Lipnitskii-2015,MierczynskiS-2013}.
However LCEs and LEs themselves are used because of their geometric meaning.

For numerical computation of LCEs and LEs there are
developed various continuous and discrete algorithms
based on QR and SVD decompositions of fundamental matrix.
However such algorithms may not work well in the case of coincidence or closeness
of two or more $\LCE_i$ or $\LE_i$
and in the case of irregular linearizations.
Also it is important to remark that numerical computation of LCEs and LEs
can be done only for a finite time $t=T$
(i.e. there are computed $\LCE^o_i(T,x_0)$ and $\LE^o_i(T,x_0)$),
justification of the choice of which is usually omitted,
while it is known that in such computations
unexpected ``jumps'' can occur (see e.g. \cite[p.116, Fig.6.3]{ChaosBook}).
Various methods (see, e.g. \cite{WolfSSV-1985,RosensteinCL-1993,AbarbanelBST-1993,HeggerKS-1999})
are also developed for the estimation of LEs from time series.
However there are known examples in which the results of such computations
differ substantially from the analytical results \cite{TempkinY-2007,AugustovaBC-2015}.
In \cite{SanderY-2015} it is noted that
these examples call
into serious question whether the Lyapunov exponents
for time series data give any sort of meaningful
quantitative measurement of the original system.

The existence of different definitions of exponential growth rate, computational methods,
and related assumptions led to the appeal:
{\it ''Whatever you call your exponents, please state clearly how are they being computed''}
\cite{ChaosBook}.

\section{Lyapunov dimension}

Various characteristics of chaotic behavior are based on LCEs and LEs.
The sum of positive exponents 
is used \cite{Millionschikov-1976,Pesin-1977}
as the characteristic of Kolmogorov-Sinai entropy rate \cite{Kolmogorov-1959,Sinai-1959}.
Another measure of chaotic behavior is the Lyapunov dimension
\cite{KaplanY-1979,Ledrappier-1981,ConstantinFT-1985,EdenFT-1991,Hunt-1996}.

Introduce the Kaplan-Yorke formula \cite{KaplanY-1979}
of the local Lyapunov dimension
with respect to the ordered set $\{\LE_i^o(t,x_0)\}_{1}^{n}$
of the finite time Lyapunov exponents.
For that, we define the largest integer $j=j(t,x_0) \in \{1,..,n-1\}$
such that
\[
 \begin{array}{c}
 \sum\limits_{i=1}^{j(t,x_0)}\LE_i^o(t,x_0) \geq 0, 
 \frac{\LE_1^o(t,x_0) + \ldots + \LE_{j(t,x_0)}^o(t,x_0)}{|\LE_{j(t,x_0)+1}^o(t,x_0)|} < 1.
 \end{array}
\]
The function $d_{\rm L}^{\rm KY}(t,x_0) = 0$ if $\LE^{o}_1(t,x_0) < 0$
and $d_{\rm L}^{\rm KY}(t,x_0)=n$ if $\sum_{i=1}^{n}\LE^{o}_i(t,x_0) \geq 0$,
otherwise
\begin{equation}\label{formula:kaplanfinite}
 d_{\rm L}^{\rm KY}(t,x_0)
 = j(t,x_0) + \cfrac{\LE_1^o(t,x_0) +
 \ldots + \LE_{j(t,x_0)}^o(t,x_0)}{|\LE_{j(t,x_0)+1}^o(t,x_0)|}.
\end{equation}
It is well known that the Lyapunov dimension is not a ``real'' dimension\footnote{
This is not a dimension in a rigorous sense (see, e.g. \cite{HurewiczW-1941,Kuratowski-1966});
e.g., in \cite[Fig.~7;p.1439]{LeonovKM-2015-EPJST} a local B-attractor,
which includes equilibria and separatrices, has $\dim_L \approx 2.8$.},
however, for an attractor $K$ the Kaplan-Yorke formula
of the local Lyapunov dimension
with respect to the finite time Lyapunov exponents
gives an upper bound for the Hausdorff dimension
(it follows from \cite{DouadyO-1980}):
\[
  \dim_{\rm H}K \leq \sup_{x \in K} d_{\rm L}^{\rm KY}(t,x).
\]

While in numerical experiments
we can consider only finite time $t$ and \eqref{formula:kaplanfinite},
from a theoretical point of view,
it is interesting to study the limit behavior
and similarly introduce
the Kaplan-Yorke formula with respect to the Lyapunov exponents:
$d_{\rm L}^{\rm KY}(x_0)=d_{\rm L}^{\rm KY}\big(\{\LE^o_i(x_0)\}_{1}^{n}\big)$
(see, e.g. \cite[p.33,\,eq.3.37]{ConstantinFT-1985},\cite{EdenFT-1991}).

The properties of Lyapunov dimension of attractors of dynamical systems
are considered in details
in the books \cite{Pesin-1988,Temam-1997,BoichenkoLR-2005}
(see also the recent surveys \cite{BarreiraG-2011,Leonov-2012-PMM}).

The idea of construction of \eqref{formula:kaplanfinite}
may be used with other types of Lyapunov exponents.
Thus other definitions for the Lyapunov dimension can be used
for the construction of \eqref{formula:kaplanfinite},
but care shall be taken to establish it relation
with the Hausdorff dimension.

Relying on ergodicity,
the Lyapunov dimension of attractor is often computed along one trajectory,
which is attracted or belongs to the attractor.
But, in general,
one has to consider a grid of points on the attractor and
the corresponding local Lyapunov dimensions \cite{KuznetsovMV-2014-CNSNS}.
Note, for example, that the Lyapunov dimension of the global Lorenz attractor
coincides with its value in the zero saddle point
and is larger then the Lyapunov dimension of the classical local Lorenz attractor
\cite{EdenFT-1991,DoeringG-1995,Leonov-2002,LeonovKKK-2015-arXiv-Lorenz}.

Along with commonly used numerical methods for estimating and computing the Lyapunov dimension,
there is an analytical approach,
based on the direct Lyapunov method using Lyapunov-like functions, which was proposed by Leonov
\cite{Leonov-1991-Vest,Leonov-2002,LeonovK-2015-AMC}.
This method requires the consideration of various changes of variables.
Thus, the question arises whether LEs
and related characteristics are invariant under such changes
(see, e.g., \cite{OttWY-1984}
{\it ''Is the Dimension of Chaotic Attractors Invariant under Coordinate Changes?''}
and \cite{DettmannFC-1995,SprottHH-2014}).
Also for the correctness of definition of the Lyapunov dimension
it is necessary to show that the definition is independent of the choice of fundamental matrix
of linearized system.

For LCEs such analysis is due to Lyapunov:
he introduced a notion of \emph{normal fundamental matrix},
whose sum of LCEs is less or equal to
the sum of LCEs of any other fundamental matrix.
Since the columns with different LCEs are linearly independent,
a linear system can have no more then $n$ different values of LCEs.
For any fundamental matrix $X(t,x_0)$
there exists a non-singular linear transformation $Q$
such that the fundamental matrix $X(t,x_0)Q$ is a normal fundamental matrix.
Note that all fundamental matrices have the same
largest LCE, which coincides with the largest LE.
For regular linearizations the set of LCEs  of normal fundamental matrix
coincides with the set of LEs.
But in the general case the set of LCEs of a normal fundamental matrix
may not be equal to the set of LEs.
Also Lyapunov showed that the so-called Lyapunov transformation
$x=L(t)y$
(non-degenerate linear transformations $L(t): {\Bbb R}\,^n\,\rightarrow\,{\Bbb R}\,^n$
such that $L(t), L^{-1}(t),\dot L(t)$ are bounded and continuous)
of coordinates of linear systems \eqref{sfl}
preserves LCEs of this linear system.
In particular, a corollary of his consideration is that
the diffeomorphism $y = D(z)$ of the phase space
does not change $\LCE$s of the bounded trajectories
of nonlinear system \eqref{sf}
(see, e.g. recent discussion in \cite{EichhornLH-2001}).


\section{Invariance of LEs under the change of coordinates}

Next it is rigorously shown that $\LE$s of linearized system \eqref{sfl} are independent of
the choice of fundamental matrix
and invariant under diffeomorphism of the phase space.

Suppose that $X(t,x_0)$ is a fundamental matrix of linear system \eqref{sfl}
and all its LEs are finite.
Consider a nondegenerate matrix $Q$ ($\det Q \neq 0$),
and suppose $\widetilde{X}(t,x_0)=X(t,x_0)Q$.

\begin{proposition}
 \[
   \lim\limits_{t \to +\infty} \bigg(\LE^{o}_i(X(t,x_0)) - \LE^{o}_i(\widetilde{X}(t,x_0)) \bigg) = 0,
   \quad i=1,..,n.
 \]
\end{proposition}
 {\bf Proof.} Consider the sets of singular values of matrices $X(t,x_0)$, $\widetilde{X}(t,x_0)$, $Q$,
 and $Q^{-1}$ in descending order
 (all singular values is strictly greater 0 since
 the matrices are nonsingular):
 \[
 \begin{aligned}
 &\sigma^{o}_1(Q)\geq...\geq\sigma^{o}_n(Q)>0,
 \
 \sigma^{o}_1(Q^{-1})\geq...\geq\sigma^{o}_n(Q^{-1})>0,
 \\ &
 \sigma^{o}_1(X(t,x_0))\geq...\geq\sigma^{o}_n(X(t,x_0))>0,
 \\ &
 \sigma^{o}_1(\widetilde{X}(t,x_0))\geq...\geq\sigma^{o}_n(\widetilde{X}(t,x_0))>0.
 \end{aligned}
 \]

 By the Horn inequality for singular values
 (see \cite{HornJ-1994-book}) for $\widetilde{X}(t,x_0)=X(t,x_0){Q}$
 and $X(t,x_0)=\widetilde{X}(t,x_0)Q^{-1}$ one obtains
 \[
 \begin{aligned}
 & 0<\prod\limits_{i=1}^{k}\sigma^{o}_i(\widetilde{X}(t,x_0))
 \leq \prod\limits_{i=1}^{k}\sigma^{o}_i(X(t,x_0))\sigma^{o}_i(Q),
 \\ &
 0<\prod\limits_{i=1}^{k}\sigma^{o}_i(X(t,x_0))
 \leq \prod\limits_{i=1}^{k}\sigma^{o}_i(\widetilde{X}(t,x_0))\sigma^{o}_i(Q^{-1})
 \\ &
 \quad k=1,..,n.
 \end{aligned}
 \]
 Hence
 \[
 \begin{aligned}
  & 0<\prod\limits_{i=1}^{k}\sigma^{o}_i(\widetilde{X}(t,x_0))
 \leq \prod\limits_{i=1}^{k}\sigma^{o}_i(X(t,x_0))\sigma^{o}_i(Q)
 \leq \\ &
 \leq \prod\limits_{i=1}^{k}\sigma^{o}_i(\widetilde{X}(t,x_0))\sigma^{o}_i(Q^{-1})\sigma^{o}_i(Q)
 \quad k=1,..,n
 \end{aligned}
 \]
 and by \eqref{xiprop}
 \[
 \begin{aligned}
  & -\infty < -\sum\limits_{i=1}^{k}\LE^{o}_i(Q)
 \leq \\ &
 \qquad \leq
 \sum\limits_{i=1}^{k}\LE^{o}_i(X(t,x_0))
 - \sum\limits_{i=1}^{k}\LE^{o}_i(\widetilde{X}(t,x_0))
 \leq \\ &
 \leq
 \sum\limits_{i=1}^{k}\LE^{o}_i(Q^{-1})
 \quad k=1,..,n.
 \end{aligned}
 \]
 It can be found a constant $c>0$ such that
 \begin{equation}\label{LEQ}
 -\frac{c}{t}<-\sum\limits_{i=1}^{k}\LE^{o}_i(Q),\quad
 \sum\limits_{i=1}^{k}\LE^{o}_i(Q^{-1}) < \frac{c}{t}
 \quad k=1,..,n.
 \end{equation}
 Then
 \[
 \begin{aligned}
 -\frac{c}{t}
 <
 \sum\limits_{i=1}^{k}\LE^{o}_i(X(t,x_0))
 -\sum\limits_{i=1}^{k}\LE^{o}_i(\widetilde{X}(t,x_0))
 <
 \frac{c}{t}, \\
 \quad k=1,..,n.
 \end{aligned}
 \]
 Since, by assumption, LEs of matrices $X(t,x_0)$ are finite,
 LEs of the matrix $\widetilde{X}(t,x_0)$ are also finite.
 Finally, one has
 \[
 \lim\limits_{t \to +\infty}
 \big(\LE^{o}_i(X(t,x_0)) - \LE^{o}_i(\widetilde{X}(t,x_0))\big)=0
 \quad i=1,..,n.
 \]
 $\blacksquare$

\noindent From the above consideration it follows that the proposition is valid also for the matrix $\widetilde{X}(t,x_0)=QX(t,x_0)$.

Consider nonlinear system \eqref{sf} under the change of coordinates $y = D(z)$,
where $D$ is a diffeomorphism.
Under the change of coordinates
the trajectory $z(t,x_0)$ is mapped to the bounded trajectory
$y(t,D(x_0))=D(z(t,x_0))$.
Then the corresponding fundamental matrix
is $W(t,D(x_0))=D'_z(z(t,x_0))X(t,x_0)(D'_z(x_0))^{-1}$
(see, e.g. \cite{EichhornLH-2001,Leonov-2008}).
Since $z(t,x_0)$ is assumed to be bounded and
$D'_z$ and $(D'_z)^{-1}$ are continuous,
$D'_z(z(t,x_0))$ and $(D'_z(z(t,x_0)))^{-1}$ are bounded in $t$
and, therefore, an estimate similar to \eqref{LEQ} occurs for their LEs.
\begin{corollary}
 \[
  \lim\limits_{t \to +\infty}\!\!\!\big(\LE^{o}_i(W(t,D(x_0))) - \LE^{o}_i(X(t,x_0)) \big)=0,
  \, i=1,..,n.
 \]
\end{corollary}

\begin{corollary}
The Kaplan-Yorke formula
of the local Lyapunov dimension
with respect to the Lyapunov exponents
$\{\limsup\limits_{t \to +\infty}\LE^{o}_i(W(t,D(x_0)))\}_1^n$
and \newline
$\{\limsup\limits_{t \to +\infty}\LE^{o}_i(X(t,x_0))\}_1^n$
gives the same value.
\end{corollary}

\section{Conclusion}
In the paper the invariance of Lyapunov exponents and the corresponding Lyapunov dimension under diffeomorphism of the phase space is shown.
Similar results can be obtained for a discrete system $z(k+1)=F(z(k))$
(see, e.g. \cite{KuznetsovL-2005,LeonovK-2007,Kuznetsov-2008}).
Note that various changes of coordinates
are widely used  nowadays for unified study of the classes
of dynamical systems
(see,  e.g., a recent paper on
differences and similarities in the analysis of the {L}orenz, {C}hen, and {L}u systems
\cite{LeonovK-2015-AMC})
and in the analytical computation of the Lyapunov dimension \cite{LeonovAK-2015}.

\bigskip
\section*{Acknowledgments}
This  work  was supported by Russian Scientific Foundation project 14-21-00041
and Saint-Petersburg State University.


\end{document}